\documentclass[usenatbib]{mn2e}
\usepackage{graphicx}

\newcommand \Ngal {\hbox{$N_{\rm gal}$}}
\newcommand \gpc       {{\rm\ Gpc}}
\newcommand \mpc       {{\rm\ Mpc}}

\newcommand \ie        {\hbox{\it i.e.,} }

\newcommand \msol      {{\rm\ M}_\odot}

\newcommand \hinv      {\hbox{$\, h^{-1}$} }

\newcommand \se        {{\!=\!}}
\newcommand \sims      {{\sim \!}}

\def\spose#1{{\hbox to 0pt{#1\hss}}}
\def\lta{{\mathrel{\spose{\lower 3pt\hbox{$\mathchar"218$}}
      \raise 2.0pt\hbox{$\mathchar"13C$}}}}
\def\gta{{\mathrel{\spose{\lower 3pt\hbox{$\mathchar"218$}}
      \raise 2.0pt\hbox{$\mathchar"13E$}}}}

\begin{document}
\title{Red Sequence Cluster Finding in the Millennium Simulation}
\author[Cohn et al]{J.D. Cohn$^{1}$,
A.E. Evrard$^{2}$,
M. White$^{3,4}$,
D. Croton$^{4}$,
E. Ellingson$^{5}$ \\
$^{1}$Space Sciences Laboratory, Univ. of California, Berkeley\\
$^{2}$Departments of Physics and Astronomy and MCTP,
   Univ. of Michigan \\
$^{3}$Department of Physics, Univ. of California, Berkeley \\
$^{4}$Department of Astronomy, Univ. of California, Berkeley \\
$^{5}$Department of Astrophysical and Planetary
Sciences, Center for Astrophysics \& Space Astronomy (CASA),
Univ. of Colorado}

\maketitle

\begin{abstract}
  We investigate halo mass selection properties of red-sequence
  cluster finders using galaxy populations of the Millennium
  Simulation (MS).  A clear red sequence exists for MS galaxies in
  massive halos at redshifts $z < 1$, and we use this knowledge to
  inform a cluster-finding algorithm applied to $500 \hinv\mpc$
  projections of the simulated volume.  At low redshift ($z=0.4$), we
  find that $90\%$ of the clusters found have galaxy membership
  dominated by a single, real-space halo, and that $10\%$ are blended
  systems for which no single halo contributes a majority of a
  cluster's membership.  At $z=1$, the fraction of blends increases to
  $22\%$, as weaker redshift evolution in observed color extends the
  comoving length probed by a fixed range of color.  Other factors
  contributing to the increased blending at high-$z$ include
  broadening of the red sequence and confusion from a larger number of
  intermediate mass halos hosting bright red galaxies of magnitude
  similar to those in higher mass halos.  Our method produces catalogs
  of cluster candidates whose halo mass selection function,
  $p(M|\Ngal,z)$, is characterized by a bimodal log-normal model with
  a dominant component that reproduces well the real-space
  distribution, and a redshift-dependent tail that is broader and
  displaced by a factor $\sim 2$ lower in mass.  We discuss
  implications for X-ray properties of optically selected clusters and
  offer ideas for improving both mock catalogs and cluster-finding in
  future surveys.
\end{abstract}

\section{Introduction}

The abundance and distribution of massive dark matter halos provide a
sensitive probe of cosmology and theories of structure formation.
The galaxies within these halos also have their evolution strongly
affected by their hosts.
Clusters of galaxies are the observational realization of such halos
which has inspired multi-wavelength campaigns to find and characterize 
them.
With the advent of large format CCD cameras on large telescopes, which
can identify galaxies to high redshifts over wide fields,
there has been renewed interest in optical searches for clusters using
multicolor imaging
\citep{Kai98,Lub00,GlaYee00,GlaYee05,Gla06,Mil05,Koe07}, see
\citet{Gal06} for a review of optical cluster finding methods.
In particular, methods which identify the cluster red sequence
\citep{BowLucEll92,LoC97,GlaYee00,LoCBarYee04,GalLubSqu05,
GlaYee05,Gla06,Wiletal06}
have attained significant success
  in identifying cluster candidates over
wide fields to $z\simeq 1$ and above.

Because red sequence galaxies dominate the cluster population, including the
reddest galaxies at a given redshift and becoming redder with increasing
redshift, the restriction to red sequence colors approximately isolates a
redshift slice. This redshift filtering increases the signal-to-noise
of cluster detection by largely eliminating projection effects from
unassociated structures along the line of sight.  However, contamination is
still expected from blue galaxies at even higher redshift than the cluster
and from galaxies near enough to the cluster to lie within the narrow,
red-sequence color region.  This residual contamination is the
focus of this work.

We are motivated by current red sequence based cluster searches, such as the
SDSS \citep{Koe07,Mil05}, in particular by those using two filters only such
as the RCS and the RCS-2 \citep{GlaYee00,Gla06} and SpaRCS \citep{Wiletal06}
\footnote{For up to date information about the RCS and SpaRCS surveys see
http://www.astro.utoronto.ca/$\sim$gladders/RCS/
and
http://spider.ipac.caltech.edu/staff/gillian/SpARCS.}.
We investigate the nature of the
cluster population selected by a two filter method applied to 
mock galaxy samples of the Millennium
Simulation (MS) \citep{Spr05,Cro06,Lem06,KitWhi07}.  
Throughout this paper, we use ``clusters'' to refer to 
objects found by the algorithm and ``halos'' to refer to the dark matter halos
identified in the simulation using the full 3D dark matter distribution.
  We use joint halo--cluster membership --- identifying the MS halos to which 
each cluster's galaxies belong --- to 
categorize the purity and completeness of the cluster population.
 (Joint halo--cluster membership is defined by
taking a cluster, found using the red sequence method below,
and then identifying the MS halos to which its galaxies
belong.)
Our cluster finder is patterned
after the scheme used in three dimensions to identify halos.  We apply a
circular overdensity algorithm, centered on bright $z$-band galaxies,
to spatial projections of the galaxy populations at the discrete 
redshifts $z =0.41$, $0.69$ and $0.99$.

An advantage of the Millennium Simulation is that it provides mock 
clusters
situated in their correct cosmological context as part of the evolving
cosmic web. Including the cosmic web is significant because
projections of superclusters, structures that tend to align along
filaments meeting at the cluster of interest, provide a major source
of confusion for cluster identification that is difficult to otherwise
model.   By having available the full 3D galaxy
and dark matter distribution in the simulation we are able to monitor 
and
isolate different physical effects which can influence red sequence
cluster finding.

The outline of the paper is as follows.  We describe our methods in
\S\ref{sec:methods} and give our findings for the MS
in \S\ref{sec:results}.
We consider some implications and properties of the blends in
\S\ref{sec:implications} and discuss
properties causing and correlating with the blending
which might extend beyond our particular search algorithm
and simulation in \S\ref{sec:discussion}.
We conclude in \S\ref{sec:conclusions}.  The appendix compares different
purity and completeness definitions in use.

\setcounter{table}{0}
\begin{table*}
\begin{minipage}{126mm}
\caption{Changes in redshifts, colors and cuts for three boxes used.}
\begin{center}
\begin{tabular}{|c|c|c|c|c|c|c|c|c} \hline Redshift &$z_{\rm min}$& 
$z_{\rm max}$ &
intercept&slope& max RS dist. $\Delta_{\bot}$&
$\frac{d(r-z)}{d(h^{-1}{\rm Gpc})}$ low/high &
$\frac{d\, z{\rm -mag}}{d(h^{-1} {\rm Gpc})}$ low/high\\
\hline 0.41 & 0.31 & 0.51 &0.52 &0.028& 0.078& -0.72/ 1.30 &
-0.56/0.52\\
\hline 0.69 & 0.57 & 0.81 &0.72 &0.052& 0.14& -1.10/0.36 &
-0.80/0.64\\
\hline 0.99 & 0.85 & 1.14 &0.75 &0.060& 0.18& -0.40/0.72 &
-1.08/1.40 \\
\hline \\
\label{tab:zchange}
\end{tabular}
\end{center}
Change in redshift across the
Millennium box at different redshifts, red sequence intercept and slope,
maximum distance from red sequence in color-magnitude space ($\Delta_{\bot}$),
 the  $r-z$ color 
change
across the box (to front, and then to back, per $h^{-1}$ Gpc),
and the $z$ magnitude change across the box. Color and
magnitude changes are taken from the Bruzual-Charlot (2003) model as
described in the text, see also Fig. 2a.
\end{minipage}
\end{table*}

\section{Methods} \label{sec:methods}

The context for our study is the model of the
spatial distribution of massive halos and the galaxies that inhabit them
provided by the Millennium simulation \citep{Spr05,Lem06}.
This is a collisionless
dark matter simulation performed in a periodic cube $500\,h^{-1}$Mpc 
(comoving)
on a side, using $10^{10}$ particles for a cosmology with parameters
$(\Omega_m,\Omega_\Lambda,\sigma_8,\Omega_b,h,n) =
(0.25,0.75,0.9,0.045,0.73,1.0)$.
Mock galaxies, with luminosities and colors, are generated by
post-processing the dark matter halo merger trees with a semi-analytic
prescription for the gas dynamics and feedback.  For details, see
\citet{Cro06,KitWhi07}.  In particular,
our version is that described in detail in \citet{Cro06}, however with 
the updated dust prescription of \citet{KitWhi07} which better models
dust extinction at higher redshifts.

%redc2
\begin{figure}
\begin{center}
\resizebox{3.0in}{!}{\includegraphics{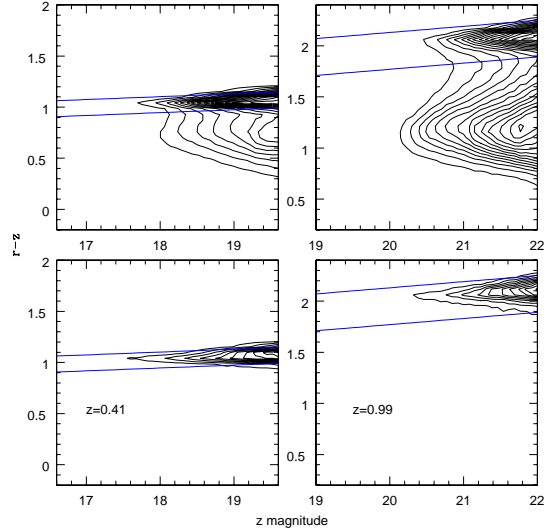}}
\end{center}
\caption{Distributions of $r-z$ colors and
magnitudes at $z=0.41$ (left) and
   $0.99$ (right) for all $z$-band magnitude-limited galaxies (top) and
   for those galaxies in halos with at least eight members (bottom).
   Contours are in steps of $\sim 770$ (left, top),
    $\sim 260$ (left, bottom),  $ \sim 360$ (right, top)
    and $\sim 60 $ (right, bottom) galaxies.
   Straight lines show the color--magnitude 
   region defining the red sequence at each redshift. 
   }
\label{fig:redseq}
\end{figure}

\begin{figure*}
\begin{center}
\resizebox{3.0in}{!}{\includegraphics{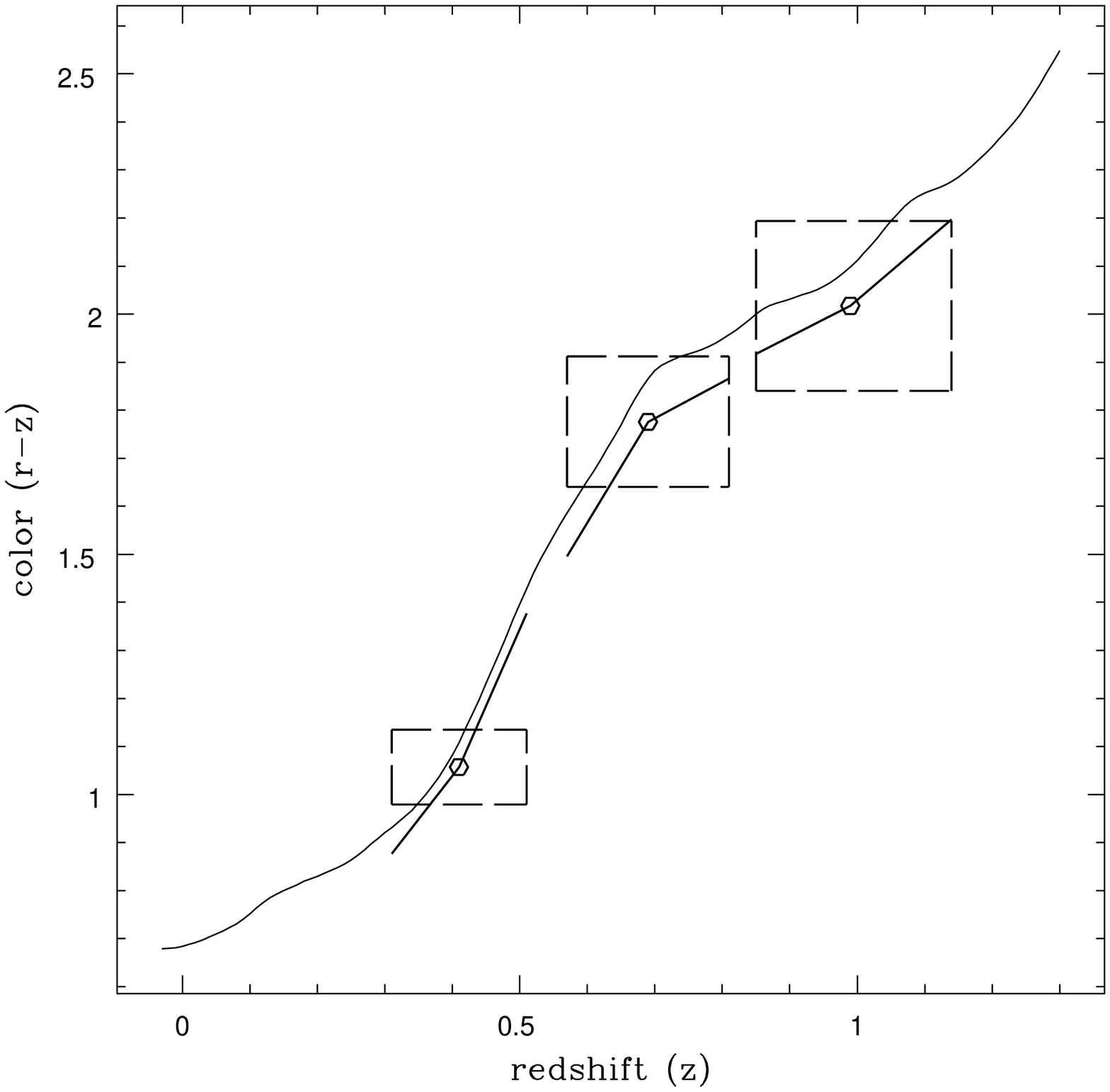}}
\resizebox{3.0in}{!}{\includegraphics{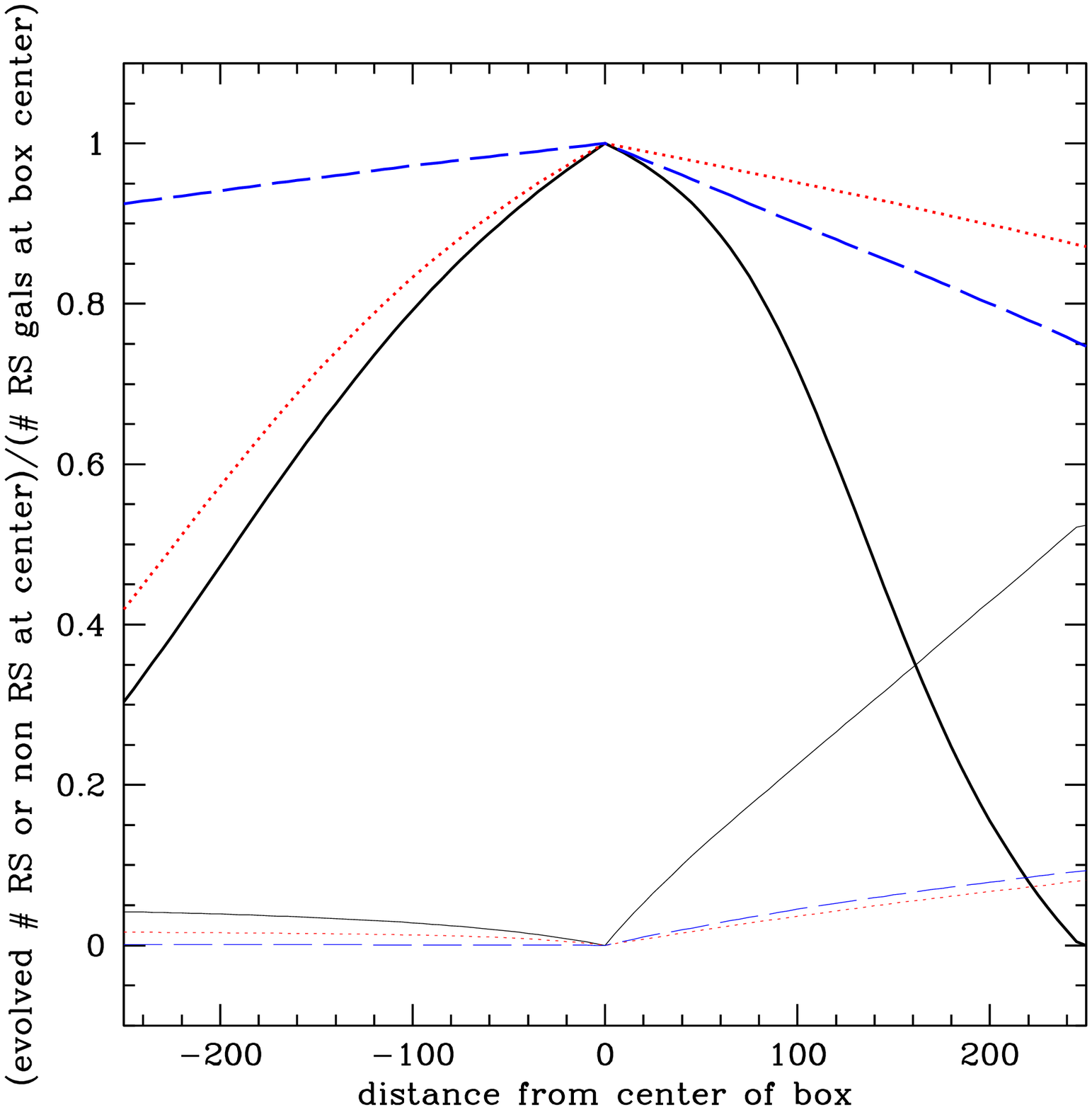}}
\end{center}
\caption{ {\bf a)} Solid lines show the $r-z$ color evolution of a $z=3$
burst population synthesis model of \protect\cite{BruCha03}.
  Circles show the mean colors of MS red sequence galaxies at the
  three redshifts we investigate, while straight line segments give
  the color gradients applied when projecting the galaxy population
  along a line-of-sight (Table~1).  Vertical portions of the dashed
  boxes at each epoch mark the foreground and background redshifts of
  the $\pm 250 \hinv\mpc$ volume, while the horizontal lines mark the
  approximate width of the red sequence.  {\bf b)} The relative
  fraction of galaxies remaining on the red sequence as a function of
  projected distance (heavy lines).  Solid, dotted, and dashed lines
  correspond to $z=0.41$, 0.69 and 0.99, respectively.  Thin lines
  give the relative number of galaxies that move onto the red sequence
  as their observed color and magnitude vary due to their
  line-of-sight displacement.  All counts are normalized by the number
  of red sequence galaxies within the volume at each redshift.  }
\label{fig:colorchange}
\end{figure*}

We focus our cluster finding investigation on local confusion,
projections on spatial scales  $\lta 250 \hinv\mpc$
of a target halo that
will, at these redshifts, be barely resolved by photometric redshifts of
the next-generation surveys 
(DES\footnote{http://www.darkenergysurvey.org},
CFHT-LS\footnote{http://cadcwww.hia.nrc.ca/cfht/cfhtls/},
Pan-Starrs1\footnote{http://pan-starrs.ifa.hawaii.edu},
KIDS\footnote{ttp://www.astro-wise.org/projects/KIDS/},
SNAP\footnote{http://snap.lbl.gov}, LSST\footnote{http://www.lsst.org}).
(Although the scales these surveys might resolve are comparable to
the box size considered here, these surveys are not necessarily using only 
the RS method described in this paper.)
We use the simulated galaxy and halo catalogues at
three fixed epochs given by redshifts $z=0.41$, $0.69$ and $0.99$.
These values span much of the expected redshift range of
interest for a survey such as the RCS.
Halos in the simulation are found by using a friends-of-friends algorithm
\citep{DEFW} and galaxy membership is determined based on this.  
The friends-of-friends linking length (0.2 times the mean interparticle spacing) can link objects into one large 
halo which by eye look to be smaller components, we note below where 
our results show signs of this effect.  
Halo masses are given in terms of $M_{200c}$ (denoted as $M$ 
henceforth), the mass enclosed within a
radius interior to which the mean density is 200 times the critical 
density
at that epoch.  At our redshifts there were 1268, 805 and 426 halos 
with
$M \geq 10^{14}\,h^{-1}M_\odot$ and 113, 47 and 19 halos with
$M \geq 3\times 10^{14}\,h^{-1}M_\odot$.

For the red sequence search, the SDSS
$r$ and $z$ filters, which bracket the 4000 Angstrom break for
approximately $0.5 \le z
\le 1 $, are employed.
At the highest redshift, we also considered $i$ band, our results for
this case are described at the end of \S 4; results below will
be given for $r-z$ unless stated otherwise.

\subsection{Galaxy Colors in Massive Halos}\label{sec:MShalos}

Fig.~\ref{fig:redseq} shows that a red sequence in $r-z$ 
vs.~$z$ exists in rich MS halos over the range of redshifts probed.
We use galaxies above $\sim \frac{1}{2}L_*$, corresponding 
to $z$-magnitudes of $19.6$, 21 and 22 at redshifts
$0.41$, $0.69$ and $0.99$, and yielding samples of  942313, 1005469 and
1054711 galaxies, respectively.  The top panels show
contours of the full, magnitude-limited 
population while lower panels show the color-magnitude behavior of galaxies
in halos with 8 or more members.  

Taking galaxies within the inner $0.5\,h^{-1}$Mpc of the centers of
the latter sample, we fit a linear relation in the $r-z$ vs.~$z$
plane.  Following \citet{Gla98}, we throw out $3\sigma$ outliers and
iterate the fit to find the slope and intercept of the red sequence.
The width of the red sequence is set to enclose 90\% of the full massive
halo galaxy sample.  The distance, $\Delta_{\bot}$, is taken
perpendicular to the red sequence line in the color-magnitude space.
\footnote{If one instead uses $>90\%$ of these galaxies, the red
  sequence widens and for high redshift slightly increases the
  contamination from projection under study here.}
Table~\ref{tab:zchange} lists the slopes, intercepts, and widths of
the red sequence for all three redshifts.  The red sequence
color-magnitude relation is a weak function of halo mass or richness,
so the parameters are not particularly sensitive to the choice of
halos with 8 or more members.

Defining the red sequence using the MS galaxy population itself means
that our color cuts are optimally tuned to the content of the MS massive
halos.  With observations, one derives color cuts using the
color-magnitude data of a target subset of galaxies, such as the
approaches used by \citet{Gla98} and \citet{Koe07}.  Comparing the
simulation results to observations, it appears that the mock red
sequence has the wrong tilt and is slightly wider than observed.  We
experimented with ``tightening'' the red sequence by moving the galaxy
colors closer to the best-fit line, but such a procedure did not have
a large effect on our conclusions so we present our results using
colors as provided.

We wish to use projections of each proper time output to create finite redshift
segments of a full sky survey.    Starting with the coeval MS galaxy samples, we
introduce passive color and magnitude evolution into spatial projections to 
mimic the behavior of a light-cone population.  The color evolution with
redshift is based on an instantaneous Bruzual-Charlot (BC) burst at $z \se 3$
 and shown for $r-z$ in Fig.~\ref{fig:colorchange}a.\footnote{We
   thank N.~Padmanabhan and B. Koester for the evolution of galaxy
   colors using \protect\citet{BruCha03} as in \citet{Pad06}.}  For
comparison, we show the average (slightly bluer) color of the MS red
sequence galaxies for our three redshifts.  The MS red sequence
galaxies are expected to be bluer than the BC model, since their stars
were not formed in a single burst at high redshift.  The MS galaxies
are also bluer than BCG's in the SDSS \citep{Ber07}.

We use this simple BC model to define piecewise constant
color gradients,  $d(r-z)/d\,{\rm redshift}$, along the line of sight, 
shown by the solid line segments in Fig.\ref{fig:colorchange}a.  
We define a $z$ magnitude gradient
analogously.    Foreground
and background color-magnitude evolution are modeled separately, with
  parameters
given in Table~\ref{tab:zchange}.   
Fainter galaxies may evolve into the $z$ magnitude cut because of
the change in observed magnitude with redshift.  To catch these 
potential
interlopers, we employ galaxy catalogues half a magnitude fainter in
$z$-band than required by the unevolved red sequence cuts.

Note that the applied color gradient
becomes progressively shallower at higher redshift.
The assumed degree of color and magnitude evolution is key since it
controls the redshift filtering power of the red sequence.
To foreshadow one of our main results,  Fig.~\ref{fig:colorchange}a
illustrates how the color evolution determines the line-of-sight path length 
probed by the red sequence color range. 
The dashed regions in Fig.~\ref{fig:colorchange}a are centered at the
average color of the red sequence galaxies at each redshift and are 
bounded
vertically by the approximate range of color of the red sequence.
They are bounded
horizontally by the redshift extents of the comoving
$\pm 250 \hinv\mpc$ sightline available within the MS volume.  At $z=0.41$,
the evolutionary color
gradients are strong enough that projected red sequence galaxies will
shift out of the target color range before the $\pm 250 \hinv\mpc$
MS boundary is reached, but this is not quite the case at
$z=0.69$ and $0.99$.

Fig.\ref{fig:colorchange}b further illustrates how the imposed color evolution
acts as a redshift filter.  Taking the color and magnitude of
each galaxy and our line of sight
gradients, Fig.\ref{fig:colorchange}b shows
the fraction of these galaxies remaining on the red sequence as a
function of line-of-sight distance.  Such galaxies will still
be potential members of a cluster centered at the origin.  A
more narrowly peaked distribution indicates a smaller
fraction of galaxies available for inclusion via projection during 
cluster finding.
As can be seen, the fraction of galaxies remaining within the red
sequence cut at large distances from the origin increases with
redshift; the red sequence selects a longer path along line of sight at
higher redshift.

The other source of contamination is galaxies that are shifted into
the red sequence by the change in observed color.  The number density of these
galaxies, normalized by the number of red sequence galaxies at the central redshift, is shown by the light lines in Fig.\ref{fig:colorchange}b.
Except for the most distant part of the box at $z=0.41$, this number
is relatively small.  Our use of a uniform color change with redshift
for all galaxies is not strictly correct for all galaxy
types.  However, blue star forming galaxies change in observed color
much more slowly with redshift than in this model, so to be shifted
erroneously into our red sequence color cut, these galaxies are
required to be at significantly higher redshift than the
cluster. Since they would then lie outside of our 500 $\hinv$ Mpc box,
they are not included in our analysis.  The strongest
contribution to interloper candidates is from galaxies which have
colors within our red sequence color cut
even though they are far from the central
galaxy along the line of sight.

\subsection{Cluster Finding Algorithm}\label{sec:algorithm}

Our algorithm defines clusters as circular regions, centered on a
bright galaxy, with red-sequence sky surface density equal to a
multiple $\Delta_p$ of the mean value at the redshift of interest.
This approach is analogous to the spherical overdensity method used to
define the halo masses.  For target centers, we work in descending
order through a list of red-sequence galaxies ranked (brightest to
dimmest) by apparent $z$-band magnitude.  This ranking is motivated by
a desire to find the rare, high mass halos first, then work down the
mass function to more common objects.

Around a potential cluster center, a radially-sorted list of red 
sequence neighbors is used to define a mean galaxy number density profile as a
function of transverse separation.  We use the periodic boundaries of the MS
to recenter the simulated volume on each candidate center.
The volume extends $250\,h^{-1}$Mpc in front and behind, and galaxy
colors are adjusted, linearly with distance in the projected direction,
as described above.  Starting with the 8 nearest neighbors, (to avoid
shot noise problems in tracing the cluster profiles at small radii),
we work outward in radius $r_{\rm gal}$ until the the number of galaxies
$\Ngal$ fails to satisfy the overdensity criterion
\begin{equation}
\Delta  \equiv \frac{\Ngal}{\bar{n} \pi r_{\rm gal}^2} \ge \Delta_p   .
\label{eqn:codefn}
\end{equation}
Here $\bar{n}$ is the mean sky surface density of red sequence
galaxies in the MS, including the effects of the applied observed
color evolution along the projected dimension.  If the overdensity criterion
is not satisfied for 8 galaxies, the object is discarded, if
$\Ngal$ meets or
exceeds a minimum of 8 galaxies, then this cluster is added to the
output list. 
All members are then recorded and removed from the
remaining list of potential cluster centers.\footnote{
Roughly the cluster will have a density of red sequence galaxies
$\Delta_p$ times the average red sequence (background) density,
$\sim 0.7/(\hinv \mpc)^2$ in our case.  The approximate change of radius
with richness can be read off from Eq.\ref{eqn:codefn}.  Note too 
that our cluster-finding algorithm traces galaxy overdensities to radii
which can potentially reach greater than 1 $\hinv$ Mpc. This algorithm
increases the survey sensitivity to truly extended structures, but may
also increase the cross-section for interlopers relative
to algorithms which search for clusters only on a limited, smaller
scale; however, a fixed aperture richness based cluster finder
performed significantly more poorly.}

Note that area overlap
of clusters is allowed, so that a single galaxy can belong to more
than one cluster (6-7\% of the galaxies end up in more than one
cluster at the lowest 2 redshifts, dropping to $4-5\%$ at higher 
redshifts; in contrast, galaxies only belong to one MS halo).
To boost statistics, we make three projections of
the simulated volume along its principal axes.  

The choice of $\Delta_p$ is discussed below.   The sensitivity of survey
purity and completeness to the choice of $\Delta_p$ is further explored in 
the appendix.

\subsection{Cluster--Halo Matching}\label{sec:match}

%mofnpts3z
\begin{figure*}
\begin{center}
\resizebox{5.0in}{!}{\includegraphics{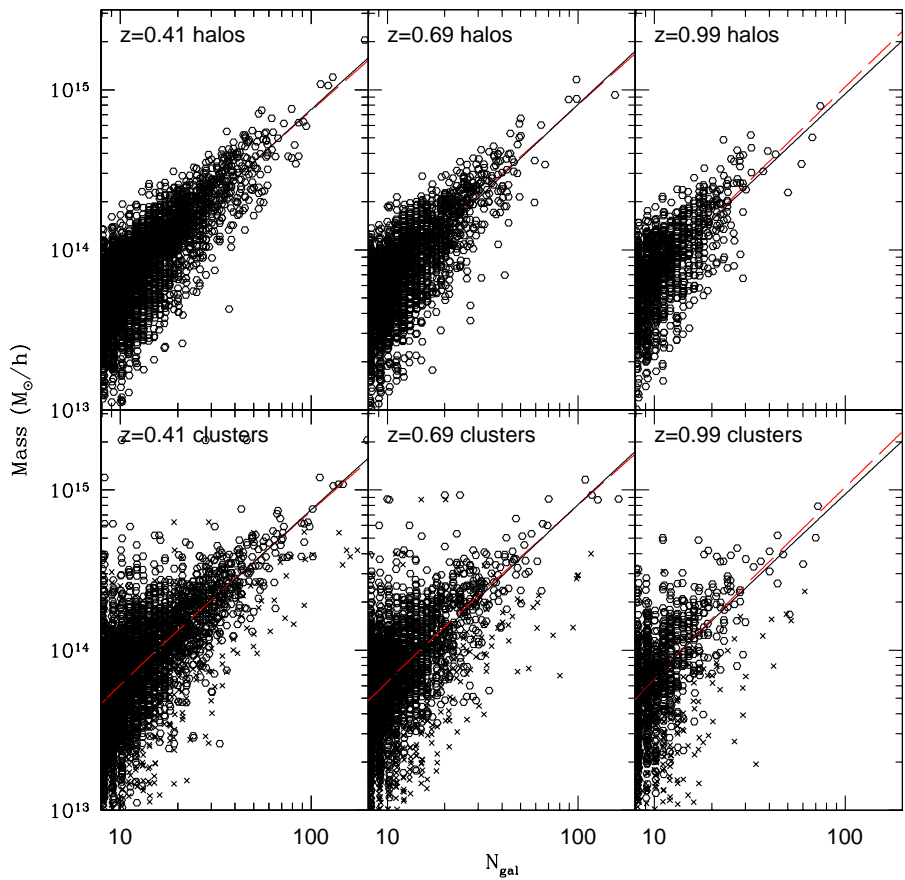}}
\end{center}
\caption{Top: Relation between halo mass and intrinsic red sequence
  galaxy richness at $z \se 0.41$, $0.69$, and $0.99$ (left to right).
  Bottom: Relation between top-ranked halo mass and cluster red
  sequence galaxy richness at the same redshifts, taken along one
  projection axis.  Crosses have $f_{1h}<0.5$ and comprise
  $(12\%,15\%,20\%)$ of the $N_{\rm gal} \geq 8$ clusters. Solid
  (dashed) lines are least-squares fits for $\Ngal \ge 8$ halos (clean
  clusters). }
\label{fig:NgalM200}
\end{figure*}

The clusters found by the search algorithm can be linked back to the 
dark
matter halos in the simulation using their respective lists
of galaxy members.  A perfect algorithm would be complete with respect
to halos and have no false positives, \ie no clusters that appear
rich on the sky but are actually several less rich systems aligned
along the line-of-sight.  In addition, the halo and cluster richnesses
would agree.  A perfect algorithm would therefore recover
the intrinsic distribution of halo mass $M$ as a function of red
sequence galaxy richness $\Ngal$.  This distribution is shown in
the top panels of Fig.~\ref{fig:NgalM200}.

At all redshifts, a mean, red sequence richness of $N_{\rm gal}\simeq 20$
above our
$z$-magnitude limit corresponds to a $\sim 10^{14}\,h^{-1}M_\odot$
halo.  When fit (throwing out 3-$\sigma$ outliers several times)
above a minimum of 8 members, we find that mass scales
with red sequence richness as $M\se M_{20} (N_{\rm gal}-1)^\alpha$,
with $\alpha=1.07$, $1.10$ and $1.10$ at $z \se 0.41$, $0.69$ and
$0.99$ respectively. The 
mass intercepts are $M_{20} \se 1.3$, $1.3$
and $1.5 \times 10^{14} \hinv\msol$ and there are $\sim$ 4100, 2900,
and 1300 $N_{\rm gal} \geq 8$ halos at these redshifts, respectively.
Note that red sequence richness is a fairly noisy tracer of mass; the
rms level of scatter is $\sim 50\%$ or higher above the richness cut
of $\Ngal \se 8$ (a detailed discussion of scatter in richness vs.
mass can be found in \citet{WhiKoc02,DaiKocMor07}).
The richness we use in finding the clusters may not be the best 
richness to
use for getting the cluster mass (e.g. galaxy counts within some 
aperture
might be useful after the clusters are found, for finding the
clusters themselves
we found a fixed aperture performed significantly worse).  Some
  observational surveys for galaxy overdensities
account for projections of foreground/background galaxies via
a statistical subtraction of the expected number of projected
galaxies, calculated from random non-cluster pointings.  Our cluster
richness estimator, $N_{\rm gal}$, does not include such a correction;
our overdensity requirement means that approximately
$1/\Delta_p$ of the galaxies are from the background.

%getclusnums,getscatfrac

For each cluster identified in projection, we list all halos
contributing one or more of its member galaxies.
The quality of the cluster detection is measured by the top-ranked 
matched
fraction, $f_{1h}$, defined as the fraction of cluster members coming
from the halo that contributes the plurality of the cluster's red 
sequence
galaxies.  We define two classes, {\sl clean} and {\sl blended},
based on whether the plurality is or is not the majority of the 
cluster's
membership,
\begin{eqnarray}
{\rm clean} & \ : \  f_{1h} \ge 0.5, \\
{\rm blended} & \ : \  f_{1h} < 0.5,
\end{eqnarray}
We assign to each cluster the mass of its top-ranked halo found through
member-matching.  If two (or more) halos contribute the same number of
  galaxies, and are both top-ranked, we
take the most massive.

\section{Results} \label{sec:results}

An ideal cluster catalog would be {\sl pure}, {\sl complete} and {\sl
   unbiased} with respect to halos.  A perfectly {\sl pure} sample
would have no accidental projections; all the galaxies of any chosen
cluster would be common members of a single, dark matter halo.  A
perfectly {\sl complete} sample would be one for which each halo in
the survey volume appears once, and only once, in the list of
clusters.  Finally, an {\sl unbiased} cluster catalog would contain
clusters that reproduce the mean mass-richness relation defined by
halos.  In this section, we consider these issues, both in the context
of setting our circular overdensity threshold and in the results
obtained.  We will see that high levels of purity and completeness are
achieved, and that the cluster samples are nearly unbiased.  (Many
definitions of purity and completeness exist in the literature, we
describe and compare several of them in the appendix, and detail our
definitions as we use them below.)

%ffound7
\begin{figure*}
\begin{center}
\resizebox{5.1in}{!}{\includegraphics{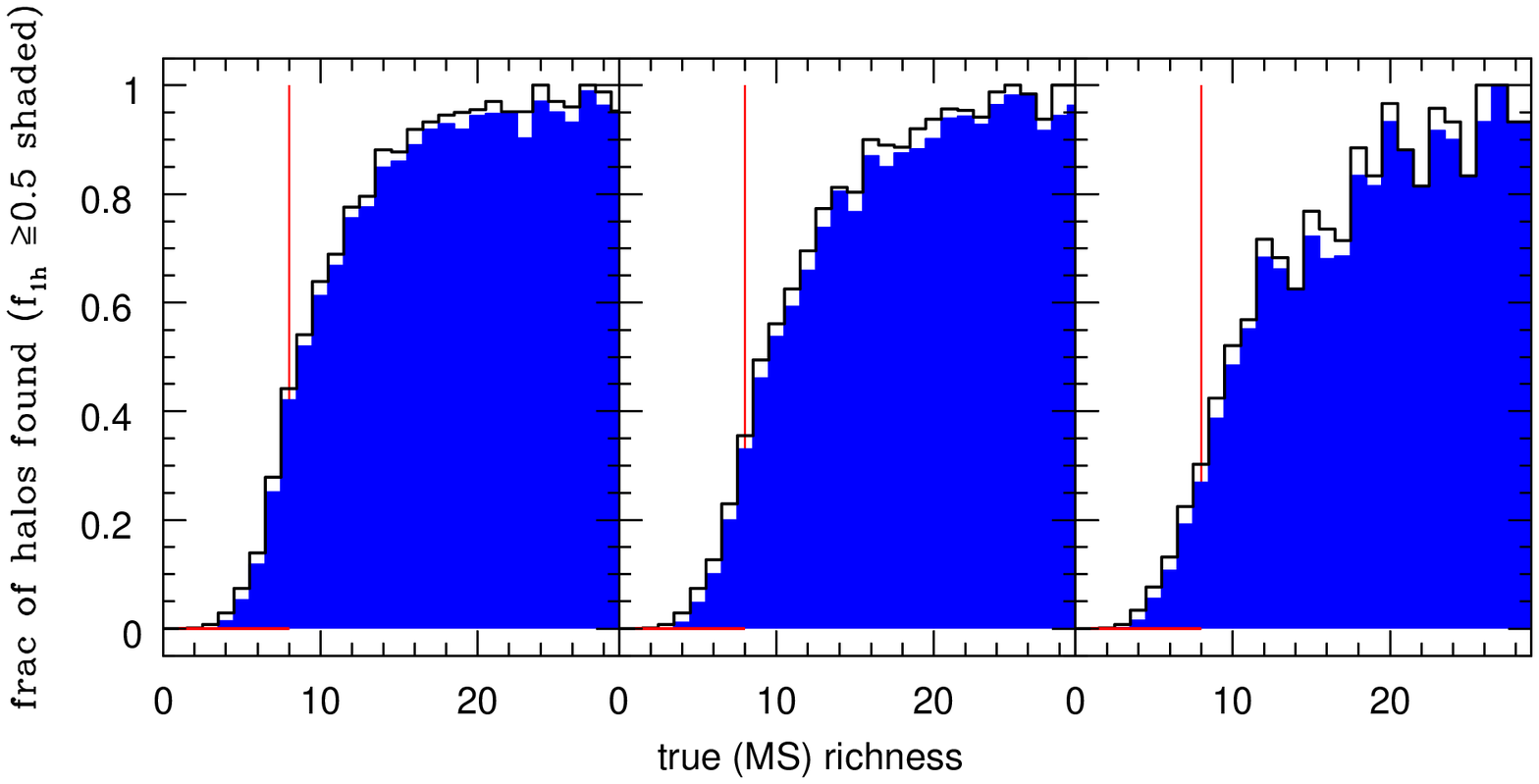}}
\end{center}
\caption{Differential completeness of the $N_{\rm obs} \ge 8$ cluster
  population with respect to halos as a function of their intrinsic
  red-sequence richness.  The circular overdensity defining the
  cluster population is $\Delta_p=7$ and panels show results for (left
  to right) redshifts $z=0.41$, $0.69$, $0.99$.  Here, completeness
  is the fraction of halos that contribute the plurality of a
  cluster's red sequence galaxy population.  The solid line is the
  fraction associated with all clusters and the shaded region is
  fraction found in clean ($f_{1h}\geq 0.5$) clusters.  The vertical
  line is the minimum imposed cluster richness imposed ($N_{\rm min}=8$).
  Projection effects introduce scatter between intrinsic and
  apparent richness that blurs the sharp observed threshold into a
  smooth intrinsic selection function.  }
\label{fig:ffound}
\end{figure*}
%countscall
\begin{figure*}
\begin{center}
\resizebox{5.1in}{!}{\includegraphics{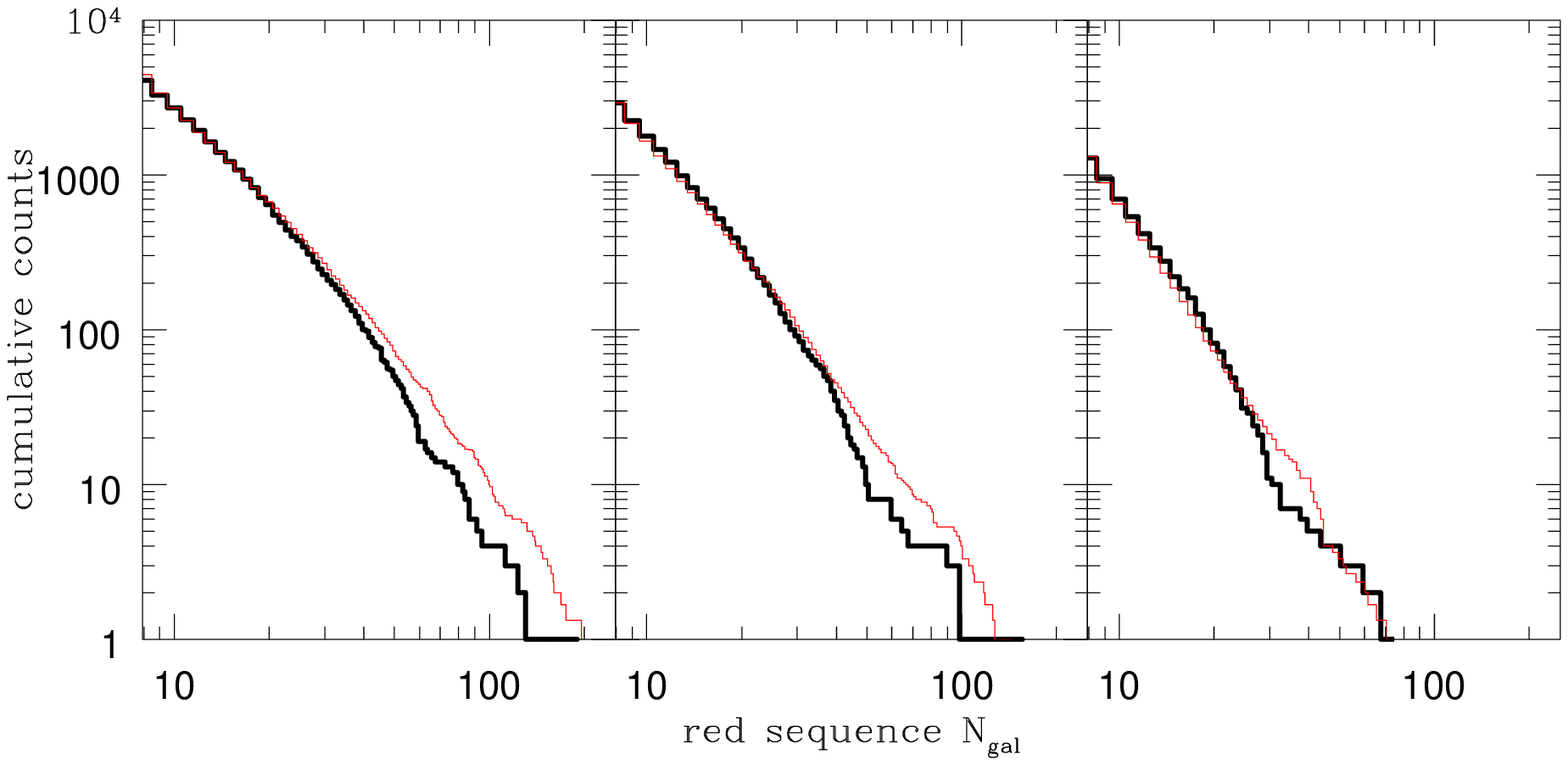}}
\end{center}
\caption{Cumulative number of halos as a function of red sequence
richness $\Ngal$ (bold) compared to the projection-averaged number 
of clusters
found with the circular overdensity algorithm with $\Delta_p=7$ (light) at 
redshifts $z=0.41$, $0.69$ and $0.99$ (left to right).}
\label{fig:nofn}
\end{figure*}
\subsection{Cluster finder threshold and two examples 
}\label{sec:tuning}

The cluster catalogs produced by the search algorithm depend on the
value of the number density threshold $\Delta_p$.  Choosing too high a
value will pick out only the cores of the richest halos, resulting in a
catalog that is pure and complete at very high masses, but is otherwise
incomplete.  Picking too low a value will extend the search into the
periphery of halos, leading to a catalog that, although complete across 
a wide range of masses, suffers from impurities due to blending multiple
halos into a single cluster.  

Our choice of $\Delta_p=7$ and $N_{\rm obs} \geq 8$ for clusters provides 
samples that are highly complete for $N_{\rm true} \geq 20$ halos.  
Fig.~\ref{fig:ffound} shows a measure of
completeness, the fraction of halos assigned as top-ranged matches to
clusters with $N_{\rm gal} \geq 8$.  The completeness is very high for
halos with intrinsic $N_{\rm true} \geq 20$, but it drops considerably
for lower-richness halos.  More halos are missed at higher redshift,
and these tend to have extended, filamentary shapes suggestive of
recent (or imminent) merging.  At higher redshift, the major merger
rate increases, leading to a higher fraction of disturbed halos.

Keeping the cluster richness fixed at $N_{\rm obs} \geq 8$ in order to
define whether a halo is found or not (completeness), 
samples derived with higher values of $\Delta_p$ will be more pure (have fewer 
blends) but less complete, and vice-versa for samples constructed with lower 
$\Delta_p$.  Further quantitative discussion on purity and completeness can be 
found in the appendix.

Fig.\ref{fig:nofn} shows that, at each redshift, the value
$\Delta_p=7$ produces a cluster catalog with a richness function,
$n(N_{\rm gal})$, that matches well that of the underlying halo
population.  Averaging the three projections, there are $4432, 2919$
and $1321$ clusters with $N_{\rm gal}\geq 8$ at $z=0.41$, 0.69 and
0.99, respectively.  These values compare well to the MS halo counts
of 4098, 2926, 1290 for $N_{\rm gal}\geq 8$. The scatter from the
average of cluster numbers between different lines of sight is less
than a percent at $z=0.41$ and less than four percent at $z=0.99$.

%showcluster41
\begin{figure*}
\begin{center}
\resizebox{3.0in}{!}{\includegraphics{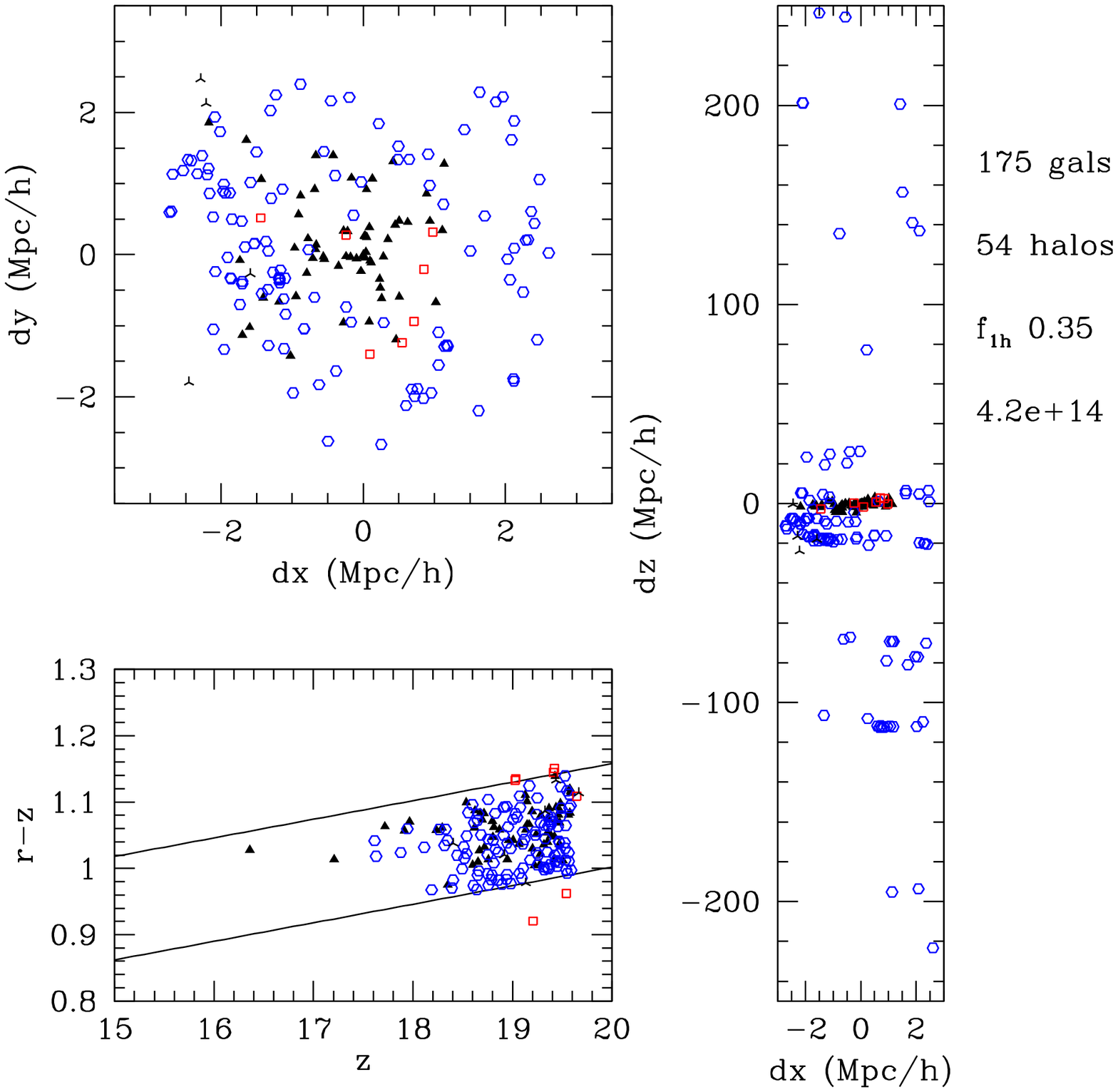}}
\resizebox{3.0in}{!}{\includegraphics{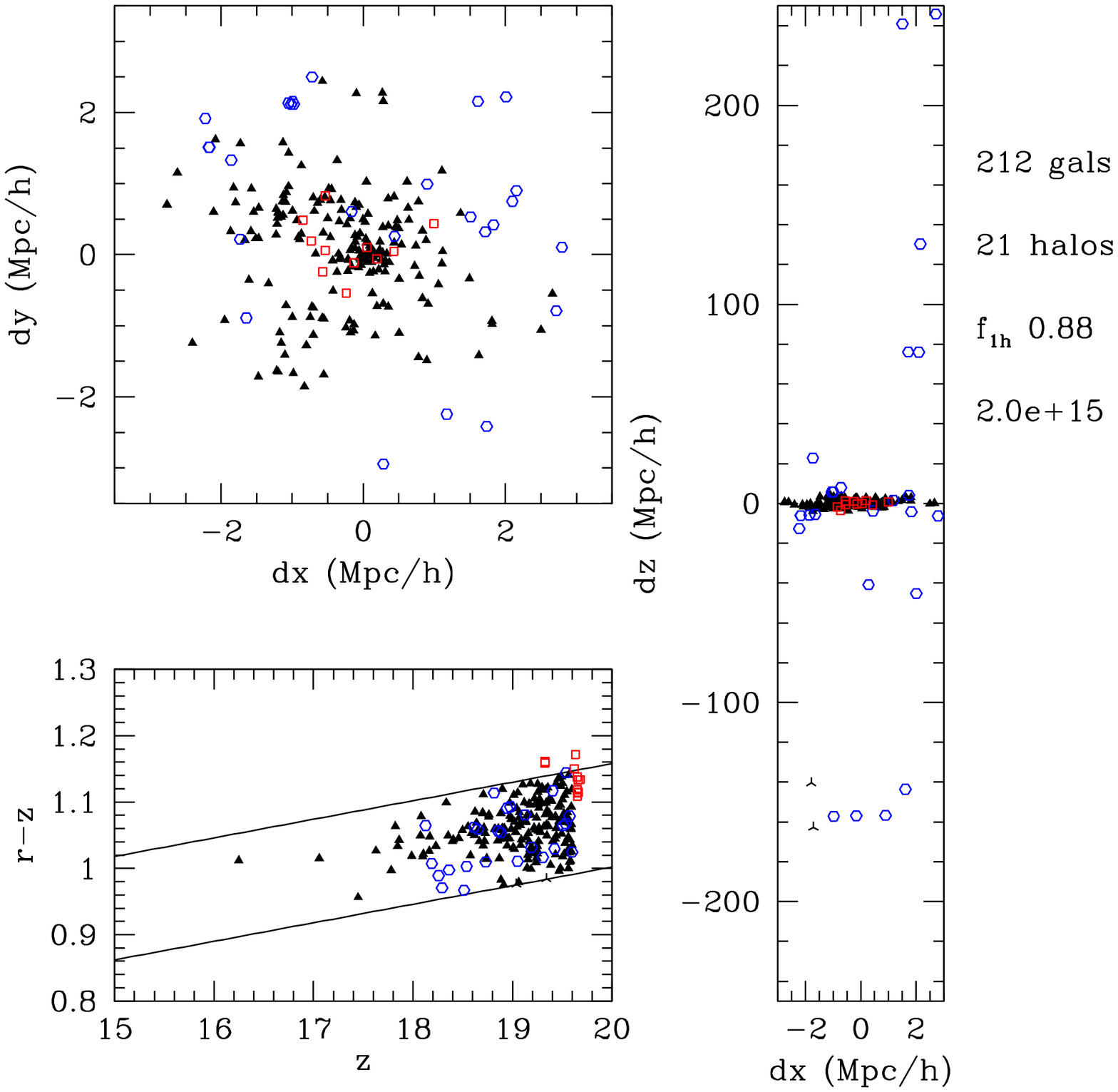}}
\end{center}
\caption{Examples of blended (left, 5th richest) and clean (right, 2nd 
richest) clusters
found at $z \se 0.41$.  Filled triangles are members of the first-rank
matched halo, open circles are other cluster members.  Open squares
are members of the best fit halo not in the cluster; they fall outside
the red sequence as can be seen in the lower left hand panel.
Triangular stars are members
of the red sequence not in the found cluster.   Comoving scales are
shown, note that the axes in the dz vs. dx figure are scaled differently.
}
\label{fig:z041c12}
\end{figure*}

The good match in number counts does not imply that the algorithm
is perfect.  In fact, the typical number of halos contributing to an
$\Ngal \ge 8$ cluster is $\sims \Ngal/4$.  
The second and fifth richest clusters found at $z=0.41$ illustrate
the range of behavior in clean and blended clusters.
Figure~\ref{fig:z041c12} shows projected positions and color-magnitude
information for sky patches centered on the two clusters.
The second richest cluster has 212 members contributed by 21 different
halos.  Members of one of the
most massive halos at that epoch, $M=2.0 \times 10^{15} h^{-1}M_\odot$,
comprise $88\%$ of the cluster members. The
remaining members come from 20 other halos, including some lying
in the foreground.  A small number of
members are contributed by halos in the background. 

The fifth richest cluster, with 175 members, presents a very
different case.  Its most massive contributing halo
has a mass $M=4.2 \times 10^{14} h^{-1}M_\odot$, which contributes 
almost
all of its own galaxies but only 35\% of the cluster's members
($f_{1h}=0.35$).
A total of 53 other halos also contribute, many lying close (within $\lta 30
\hinv\mpc$) in the foreground or background.  

Although much richer than most of the halos considered, these two 
examples
illustrate the essential projection problem that is causing the blends;
both sets of galaxies appear to be reasonable clusters in the x-y plane.
In the next two sections the statistics of the clean and blended 
clusters,
and their features, will be discussed in more detail.

\subsection{Mass selection function of clusters }\label{sec:mselect}

The mass selection function is an important ingredient for
cosmological tests with optical cluster surveys (\citet{WhiKoc02},
\citet{Roz07}).  Fig.~\ref{fig:NgalM200} (bottom) shows the relationship
between the observed richness of a cluster and the mass of its
top-ranked halo (see \S~\ref{sec:match}).  Circles show clean clusters
while small crosses show blends.  At each redshift, the clean cluster
population displays a power law mean relation remarkably similar to
that of the underyling halo population.  The slopes of the relations
agree at the few percent level; the values for halos (clean clusters)
for $N_{\rm gal}\geq 8$ are 1.07 (1.04), 1.10 (1.06), 1.10 (1.15) from low
to high redshift, respectively.  The intercepts at $N_{\rm gal}=20$
also agree at the few percent level, and could be further fine-tuned
by introducing small changes to the search threshold $\Delta_p$ at
each redshift.  At all redshifts, the circular overdensity algorithm
is effective at identifying the mean richness-mass behavior of the
underlying halo population.

\begin{figure*}
\begin{center}
\resizebox{4in}{!}{\includegraphics{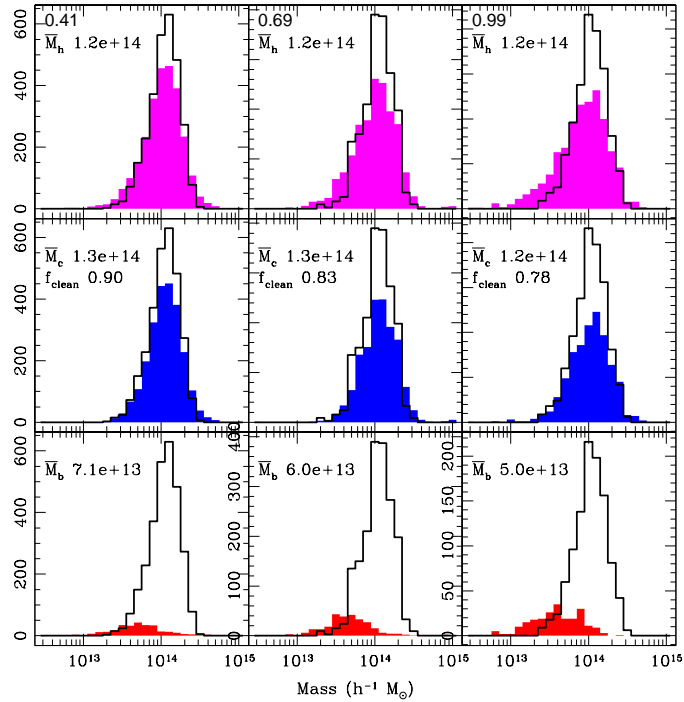}}
\end{center}
\caption{ Mass selection functions $p(M|\Ngal,z)$ with richness
  $N_{\rm gal}=(18,18,16) \pm 4$ at redshifts $z=(0.41, 0.69, 0.99)$
  (left to right).  Solid lines give the intrinsic halo mass
  distribution in these richness ranges, and are the same in each
  column.  The shaded distribution in the upper row gives $p(M|\Ngal)$
  for clusters, with $M$ the mass of its top-ranked matched halo
  (\S~2.3).  The middle row shows $p(M|\Ngal)$ for clean clusters
  ($f_{1h} \ge 0.5$) while the bottom row gives the mass
  distribution of blended clusters ($f_{1h}<0.5$).  The average
  mass of the halos/clean clusters/blended clusters are shown
  respectively in the top/middle/bottom panels for each redshift.  The
  fraction of clean clusters $f_{\rm clean}$ is also given in the middle
  row for each redshift.  }
\label{fig:mlike}
\end{figure*}
%fbest9panel1.2

The dispersion in the observed cluster sample is larger than for
halos, due to failure modes of the search algorithm.  At fixed
observed richness, blending creates a tail to low masses while
fragmentation of large halos into multiple clusters introduces a high
mass tail.  Fig.~\ref{fig:mlike} shows estimates of the conditional
halo mass distribution, $p(M|\Ngal,z)$, derived from cross-sections of
the joint likelihood data in Fig.~\ref{fig:NgalM200} in richness
ranges $\Ngal=(18,18,16) \pm 4$ at redshifts $(0.41,0.69,0.99)$,
respectively.  This choice gives a constant average halo mass, $1.2
\times 10^{14}\hinv M_\odot$, at all three redshifts.

The cluster likelihoods (shaded in the figure) are
compared with the halo distributions for the same richness ranges,
shown by solid lines.  The top row shows all
clusters, while the middle and bottom rows separate the samples into 
clean
and blended systems, respectively.  Raw counts rather than normalized
likelihoods are shown to give the number of objects.
 
At $z \se 0.41$, more than $90\%$ of clusters in the chosen richness 
range
have their dominant underlying halo contributing at least half of the 
galaxies.
The mass distribution of the found clusters matches well
the underlying halo mass likelihood.  At higher redshift, the
correspondence between halos
and clusters weakens somewhat;  the number of blends more than doubles,
from $<10\%$
at $z \se 0.41$ to $22\%$ at $z \se 0.99$.
The blended systems contribute a low mass tail to the halo mass
likelihood.  For the distributions,
the central mass of the clean clusters remains
at $1.2-1.3 \times 10^{14} \hinv M_\odot$ at all 3 redshifts,
while the central mass of the blends drops, from
$7.1 \times 10^{13} \hinv M_\odot$ at $z=0.41$
to $5.0 \times 10^{13}\hinv M_\odot$.  Thus the ratio of central masses
between the clean and blended clusters also increases with redshift.

%fbesthvalex
\begin{figure}
\begin{center}
\resizebox{3.1in}{!}{\includegraphics{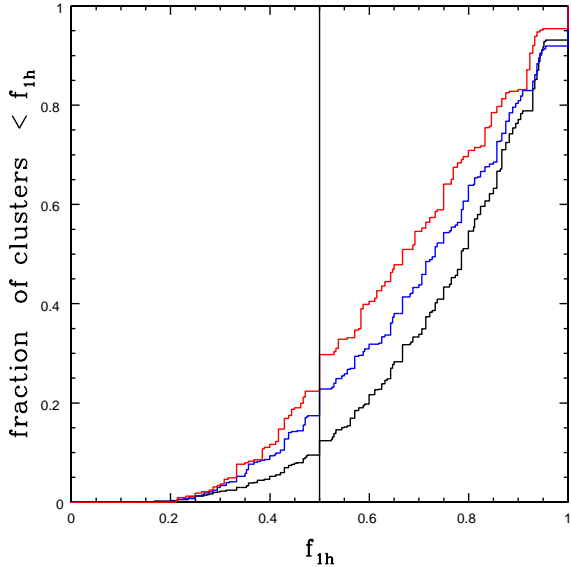}}
\end{center}
\caption{
Cumulative fraction of clusters in Fig. \ref{fig:mlike} as a function 
of their top-ranked halo overlap fraction, $f_{1h}$.  Top to 
bottom
lines are redshifts, $z \se 0.99$, $0.69$ and $0.41$.  The fraction of
galaxies from the top-ranked halo declines with increasing redshift.
The vertical line identifies the fraction of clusters that are
blends, $f_{1h} <  0.5$.
\label{fig:f1hcum}
}
\end{figure}

Our classification of clean {\sl versus\/} blended clusters is based
on a somewhat arbitrary cutoff of $0.5$ in member fraction.
Figure~\ref{fig:f1hcum} provides a more complete picture by plotting
the cumulative fraction of clusters that have top-ranked halo member
fraction $< f_{1h}$.  Here the same observed cluster richness limits as in
 Fig.~\ref{fig:mlike} are used.
Cutting at $f_{1h}\geq 0.5$, the vertical line, gives the 
clean fractions quoted in Fig.~\ref{fig:mlike}.  Analogues for other
definitions of ``clean fraction'' in terms of $f_{1h}$ can
be read off as well.  There is a clear trend with redshift, with 
clusters at
$z \se 0.99$ being less well-matched to halos than those at $z \se 
0.41$.  
The median value of $f_{1h}$
tells a similar story, decreasing from $\sims 0.8$ at $z \se 0.41$ to
$\sims 0.7$ at $z \se 0.99$.
Blending is clearly increasing at larger redshift.

Going to a higher central mass
gives similar trends, e.g.~centering on a richness corresponding to
a average $1.5 \times 10^{14} \hinv M_\odot$ halo mass at all redshifts
gives a clean fraction of 90\% at redshift 0.41 which
decreases to 76\% at redshift 0.99 for the same $\Delta_p$ as above 
($\Delta_p$
can be increased for higher richness to improve both numbers but the
increase of blends at high redshift remains).
%fbest9panel1.5

\subsection{Causes and trends for blends}

There are several effects which cause an increasing incidence of blends 
at
higher redshift.
Firstly, the change of observed color with distance is weaker,
and secondly, the red sequence is wider,
so the color-magnitude cut selects galaxies from a thicker
slice along the line of sight.
These seem to be the strongest causes and were illustrated
in Fig.\ref{fig:colorchange}.

Another way of seeing the effect of color/magnitude evolution is to
remove it entirely at $z=0.41$; the background level then increases
and the contrast between the clusters and the background declines.
Lowering $\Delta_p$ to obtain the same number of clean clusters at the
fixed mass range of Fig. \ref{fig:mlike}, we find that
  the level of blends increases to $\sim 20\%$,
very close to what is seen at $z\sim 0.99$.  Similarly, to increase
the clean fraction, one can impose the $z=0.41$ color evolution on
the $z=0.99$ population.  In this case, however, the number of
non-red sequence galaxies brought into the red sequence through our 
evolution increases strongly, limiting the degree to which blends can be 
reduced.  

A third contributing factor is that, at earlier times, the mass function 
is steeper, causing the number of possible interloper halos per target
halo (of mass $\sim 10^{14} \hinv M_\odot$, for example) to grow at 
high redshift.  The increase in intermediate-mass halos
is also enhanced because the central galaxy magnitude
is less well correlated with host halo mass at $z \se 0.99$ than at low
  redshift.
Over time, central galaxies in massive
halos grow and brighten via mergers, leading to a stronger correlation
between $z$--magnitude and halo mass.  Our cluster finding algorithm works in
descending order of luminosity.  At low redshift, the luminosity sorting 
corresponds well
to a sorting in halo mass but, at high redshift, more low mass systems are mixed
 into the range of central galaxy magnitude occupied by high mass halos.

As these factors are fairly generic, as expected,
the trend toward more blends at $z \se 0.99$ appeared in all the cases
we considered: changing definition and tightness of the red sequence, 
changing
$N_{\rm gal}$ cuts and changing the spherical overdensity requirement.
For a wide range of density cuts and modeling choices the blends have
roughly half the mass of the clean matches at $z=0.41$, and this
mass scale declines at higher redshift.

\section{Implications} \label{sec:implications}

Since blended clusters are
associated with lower mass halos, they will be evident in follow-up
studies as such.  Their mean lensing signal, X--ray luminosity and
temperature, and thermal SZ decrement should be low relative to clean
systems.  Spectroscopic signatures of substructure, in the form of
multiple peaks or other departures from Gaussianity, would also be
likely in these systems.  The imprecise centering of the multiple
components along the line-of-sight would tend to flatten the radial
number density profile.

\begin{table}
\begin{center}
\caption{Expected Cluster X-ray Properties.}
\begin{tabular}{|c|c|c|cc|} \hline
Redshift & $\langle L \rangle_{\rm halo}^a$
& $\langle L \rangle_{\rm clean}^a$ &
$\langle L \rangle_{\rm blends}^a$
& $f_{\rm blends}$ \\
\hline
0.41 & 1.4 (0.96) & 1.6 (1.0)  & 0.84 (1.5) &$0.11$ \\
0.69 & 1.4 (0.96) & 1.8 (1.1) & 0.70 (1.3) &  $0.16$ \\
0.99 & 1.8 (0.97) & 2.0 (1.1) & 0.56 (1.3) & $0.23$  \\
\hline
\end{tabular}
\label{tab:xray}
\\
$^a$ Numbers in parenthesis give the log-normal scatter, $\sigma_{\ln L}$.
\end{center}
\end{table}

Table \ref{tab:xray} provides estimates of the soft band %xraytable.dat
X-ray luminosity from our MS blended and clean clusters with richness 
18 $\pm
4$ (now fixed across redshifts), compared to values for halos of the
same richness.  We assume a power-law relation of the form
$L\propto(M/10^{14}\hinv M_\odot)^{1.6}$ \citep{Sta06}, and quote values 
normalized, arbitrarily, to the luminosity of a $10^{14} \hinv\msol$ halo
at each epoch.
We also assume scatter in the mass--luminosity relation,
$\sigma_{\ln M}=0.4$, and combine this with the dispersion in mass
for the chosen richness range (Fig.\ref{fig:mlike}) to give the
dispersion in luminosity, $\sigma_{\ln L}$.
Lower values have been suggested for
$\sigma_{\ln M}$ \citep{ReiBoe02}, but the scatter in mass at
fixed $\Ngal$ dominates the intrinsic L-M scatter anyway.

The clean clusters have mean X-ray luminosities
that tend to be slightly higher than
the corresponding values for halos of the same richness.  The
blended systems are substantially dimmer, by a factor two in the mean at
$z=0.41$, growing to a factor three at $z=0.99$.

Blends should be a generic outcome of red
sequence-based cluster finding methods, and there are indications of 
this from
initial X-ray and dynamical
observations of the RCS clusters.  In Chandra observations of 13
clusters at $0.6 < z < 1.0$, \citet{Hic05,Hic07} confirm 12
as X-ray sources at 3$-\sigma$ significance,
suggesting that  $>90\%$ of the cluster candidates are massive 
structures with
deep gravitational potential wells (see also \citet{Bli07}). However,
their X-ray luminosities were systematically lower at a given
cluster richness than seen for lower-redshift X-ray selected
clusters. Most of the clusters lay on a sequence only slightly offset
from the expected $L_x$-richness relation, but several clusters were
significantly offset. Optical spectroscopy of one of these clusters
(at $z=0.9$) showed that it consisted of several structures which are
dynamically discrete but whose red sequences were overlapping in the
survey data \citep{Gil07}-- precisely the sort of blended
system expected by the study here (see also
\citet{vanB07}).  Evidence for large scatter
between X-ray luminosity and optical richness has been seen in e.g.
\citet{YeeEll03,Gil04,LubMulPos04,Hic05,Bar06}.

Instead of using only the top-ranked halo mass to determine the
X-ray signal, we can
instead sum the luminosity of all contributing halos.   In this case,
all the cluster luminosities go up, with the clean subset increasing by 
roughly $0.3$
and the blended subset increasing by a larger amount.   Then the 
ratio of clean to blended mean luminosities changes to $\sim 1.2$ at 
low redshift and to $\sim 2.4$ at high redshift.  The 
luminosity measured by X--ray observation will depend on details of the 
projected spatial arrangement, the noise characteristics and other 
details that lie beyond the scope of this investigation.  It seems 
reasonable to consider the values quoted for the single halo case as a 
lower bound, and the values from summing all halos as an upper bound, 
on what would be observed.

\begin{figure}
\begin{center}
\resizebox{3.0in}{!}{\includegraphics{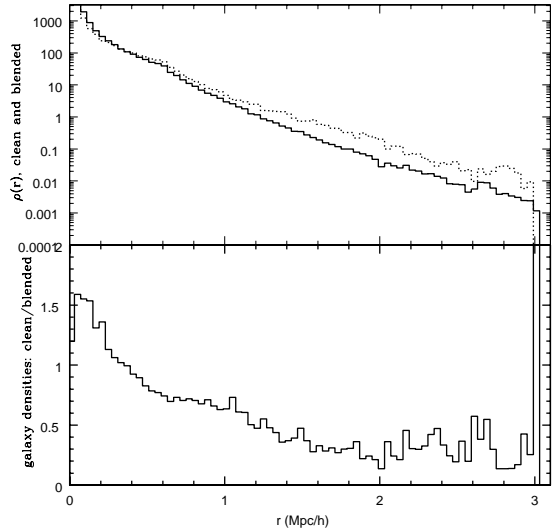}}
\end{center}
\caption{``Stacked'' profiles of clusters with $f_{1h}\geq 0.5$ divided 
by
those with $f_{1h}< 0.5$ for the $\Delta_p=7.0$  case and $z=0.41$.
The case here is representative, the trend of ratio with radius
was seen in all redshifts and color cuts. Stacking after
rescaling by the outer radius gave similar results.}
\label{fig:radial}
\end{figure}

Another difference between clean and blended systems is
in their radial cluster profiles.  Stacked profiles of the clean
and blended clusters are used to produce the density profiles,
$\rho(r)=\frac{1}{N_{\rm clus}}N(r)/(r^2 dr)$, shown in Fig. 
\ref{fig:radial}.
The clean clusters have a significantly steeper mean density
profile than the blends.  This result suggests that
a matched angular filter approach \citep{Pos96} could offer
improvements, particularly one that includes radial distance information
from photometric redshifts or colors \citep{WhiKoc02}.
Observations of colors with distance to cluster center
(e.g.~\citet{Ell01,Bli04}) and other properties (e.g.~\citet{DeL04}) are
already in place at high redshifts.
Going further down the luminosity function would provide more galaxies
to trace out the profile, but at the risk of including more faint 
background galaxies redshifted into the color region.

%fbestrdist8

The interlopers in both clean and blended clusters, as expected by
\citet{GlaYee00},
lie (slightly) more frequently in the background than the foreground.
There doesn't seem to
be a strong trend in the moment of inertia for clean versus blended
clusters; often the blends are projections, rather than objects
which are merely unrelaxed.

We also considered using $i-z$ color at high redshift, rather than 
$r-z$,
and found a similar blend fraction, even though the red sequence turns 
out
to be narrower.  This is because
the evolution of red sequence galaxies (now defined with respect to 
$i-z$)
remains very slow with redshift, thus,
as with $r-z$ color at high redshift, many of the
galaxies do not evolve out of the red sequence even when far from the 
cluster
center.  Similarly, the number of non-red sequence galaxies evolving 
into the selection window remains small across the $\pm 250 \hinv\gpc$ 
projected length.

As mentioned earlier, blends can be immediately reduced by increasing 
the
spherical overdensity criterion $\Delta_p$,
but only at the cost of losing true halos as well.
An increase in $\Delta_p$ also shifts the mass-richness
relation to lower values of $\Ngal$ compared to the intrinsic case,
and decreases the number of clusters found at fixed $\Ngal$.
These trends reflect the usual tradeoff between purity and completeness 
for cluster samples; for more discussion see, e.g.,
the appendix of \citet{WhiKoc02} and the appendix of this paper.

\section{Discussion} \label{sec:discussion}

In the above analysis, we have found properties and trends for blends 
as a function
of redshift.  Some of these results depend on particular details of the
Millennium Simulation and our method, and some are likely to be general.

Most of the increase in blends at $z \sim 1$ comes from the slower 
change of
color with increasing redshift.  This color change was not obtained directly
from the Millennium simulation but from a simple stellar population 
synthesis model that reproduces observations.  We expect this result to be 
general.  Our implementation of the color change with redshift is crude
but the candidate high redshift interlopers are mostly red
sequence galaxies, where our approximation
is best expected to hold.  As a result, we do not expect
more detailed color implementations, such as mock
light cones (e.g.~ \cite{KitWhi07} for the MS), to produce substantially
different local ($\pm 100 \hinv \mpc$) projected contamination.

The increased width of the red sequence at high redshift is
derived from the Millennium Simulation.  However, at $z \se 0.99$, the
weak color evolution combined with the deep ``green valley''
separating the red and blue populations in the MS means that our 
results are reasonably
insensitive to the precise width.  Most of the interloper galaxies are 
themselves members of the red sequence in their respective projected 
halos.   The $r-z$ color shift for $\pm 250 \hinv\mpc$ projection at
$z=0.99$ is $-0.1$ and $+0.18$, so only by compressing the red sequence 
to a width well below these values would one have an appreciable effect 
on the blended fraction.

The relative numbers of interloper halos at different redshifts is a 
property of the underlying
dark matter power spectrum and linear growth rate.  For a fixed target
mass, more interloper halos at higher redshift are expected  
generically.
Physically,
if we look at the line-of-sight distribution of the contaminating 
material
we find that the contaminating mass at large distances 
($>50\,h^{-1}$Mpc)
%fracoutdzfid
more than doubles between redshift $0.41$ and $0.99$.  This enhanced
contamination from large distances is also true on a cluster-by-cluster 
basis:
the fraction of clusters with more than e.g. 30\% of their material 
coming
from $>40 h^{-1}$ Mpc grows significantly with redshift. %outdzpics1040b
This material is far outside the cluster
virial radius and not just material which got ``caught'' by the cluster
finder before it fell in to truly be part of the cluster.
Note that superclusters of very large size have been seen out at these
redshifts, for example see recent studies of superclusters by
\citet{GalLubSqu05,Nak05}.

There are other possible interlopers as well.
For instance, adding galaxies that lie outside the MS volume will only 
increase
the amount of blended contamination.  Also, at faint magnitudes,
the increasing numbers of background blue galaxies available to
redshift into the red sequence are a potential cause for concern;
increasing numbers of blue galaxies at high redshift are observed
(e.g. \citet{Ell97}). (It should be noted that there are observational
techniques to take many of them out which we do not include here).
   We saw only a small fraction of candidate interlopers from galaxies 
outside the red sequence, except at low redshift, where the green valley is 
highly compressed (see Fig.\ref{fig:colorchange}).  This is good, as the 
observed
color and magnitude evolution of these galaxies was approximated to be 
the
same as for red sequence galaxies; we expect the interloper numbers due
to these objects is minimized at high
redshift because of their small contribution in our approximation.
In fact, taking out the observed color and
magnitude evolution of the non-red-sequence galaxies entirely (and
adjusting $\Delta_p$ to get the same value of $\bar{n} \Delta_p$)
gives similar clean fractions as in our fiducial model.

To extend our analysis of the MS
(tuning the cluster finder to cluster color profiles in more detail,
for example) requires further developments.  The MS utilizes 
sophisticated
physical models for
properties such as star formation histories, initial mass function and
stellar population synthesis models, dust production and radiative 
transfer
models, and the sensitivity of all of these to local conditions.
The resulting MS catalogues match observations of large
numbers of properties (e.g. \citet{Spr05,Lem06,Cro06,KitWhi07}) at many 
redshifts.  A detailed, multi-color comparison to observed galaxy number counts is given in \citet{KitWhi07}.  Some 
departures from observations are noted there, in particular the
over-prediction of the abundance of moderately massive galaxies 
at high redshifts, notably $z>1$.

For our cluster finding, the only properties used are
the galaxy locations and their fluxes in two filters.  Tuning the cluster
finder to more specific properties of these fluxes (for example, their radial
trends within clusters) in the MS will require higher fidelity 
galaxy formation models. 
Work is in progress to improve the model's match to observations.  
For example, known
issues in the MS under study include, for $z=0$ clusters, a
faint red satellite excess and an excess tail of bright blue
objects (but with overall blue satellite
fractions too low), and no ``green valley''\citep{Wei06a,DeL06,DeL07}.
We find a red sequence with the wrong sign for the color-magitude slope
(the brightest galaxies tend to be slightly bluer than the fainter,
see Fig.~\ref{fig:redseq}), and similarly
the blue fraction increases towards brighter luminosity
and has the wrong radial evolution within clusters
for our three redshifts.  In addition the simulation was run with the
earlier WMAP parameters rather than the WMAP3 \citep{WMAP} current
best fit cosmology.

Future improvements in optical cluster finding will require simulated
catalogues that are in better quantitative agreement with the growing
body of deep galaxy surveys.  To refine and use more sophisticated
color finders does not necessarily require all the physics employed in
the MS, which aims to explain a multitude of observational properties
rather than simply reproduce them.  Purely statistical mock catalogues
can be built on halo model fits tuned empirically to data in the
redshift range of interest.  The catalogues can focus narrowly on
observational properties relevant to the search algorithm.  In
particular, critical to a quantitative prediction of the amount of
contamination in color-selected surveys are accurate colors for
galaxies in groups and filaments in the outskirts of clusters, as
red-sequence galaxies in these regions are the most likely source of
interlopers.

Such survey-specific catalogues are crucial to understand selection
functions, in part because they allowing search algorithms to be tuned to the
cluster ``color  footprint'' and spatial profile.  This approach has already
been profitably used at low redshifts
\citep[e.g.][]{Koc03,Eke04,YanWhiCoi04,Yan05,Mil05,Wei06b,Mil05,Koe07}.
At high redshift, data sets large enough to tune such catalogues are just
coming into being; combined with modeling improvements in recent years the
construction of such catalogues is now a feasible task.

However, without such a catalogue in hand, our primary effect is
still simple to illustrate.  This effect is
that the spatial cut provided by the observed color and magnitude
cut widens as redshift increases.  Conversely a narrow spatial cut
reduces the blends strongly.  For example, taking an exact spatial cut 
for
the MS, boxes  $100\,h^{-1}$Mpc wide at all three redshifts, the clean 
fraction
becomes almost 100\% at low redshift and 95\% at high redshift. 
(Presumably the
remaining blends are due to the other contributing factors mentioned 
above.)
A slice this thick would corresponds to
a fine redshift selection, $\Delta z=0.06 (0.04)$ at
redshift 0.99 (0.41).  This level of accuracy is potentially
attainable with next-generation photometric redshifts. 

\section{Conclusions} \label{sec:conclusions}

With the advent of wide field imagers, optical searches have become a
powerful way to compile large samples of high redshift clusters.
Key to these techniques is the use of multi-color information to reduce 
the
line-of-sight contamination that plagued
earlier, single filter,
observations \citep{Abe58,Dal92,Lum92,Whi99}.
Two-filter information provides only limited redshift filtering, and
this paper begins to explore the questions of what types of objects are 
selected
by such techniques, and how this selection evolves with redshift.

We use a simple circular overdensity search algorithm on local sky
projections of the galaxy population of the Millennium Simulation,
tuned using knowledge of the red sequence present in simulated halos with
eight or more galaxies brighter than $L_\ast/2$ in the $z$-band.  The
free parameter, the density contrast $\Delta_p$, is tuned to maximize
both purity and completeness, and the choice $\Delta_p=7$ produces a
number of clusters as a function of galaxy richness that is close to
the underlying richness function of halos.

We find that essentially all clusters have some degree of projected
contamination; a cluster of optical richness $\Ngal$ typically has
red sequence members from $\Ngal/4$ halos along the line-of-sight.  In 
the
large majority of cases, the contamination is not dominant, and
most of a cluster's members are associated with a single, massive
halo.  A minority are highly blended cases in which projected
contamination is dominant, and no single halo contributes a majority
of the cluster's members.

We find an increased fraction of blends with redshift.  Although 
several factors
contribute, the most important factor appears to be weaker evolution
in the observed color of red sequence galaxies with increasing redshift.  This
effectively increases the path length searched by the
red sequence color cut, leading to a larger cross section for
accidental, line-of-sight projections.  In addition, at higher redshift,
the number of $\sim 3 \times 10^{13} \hinv M_\odot$ halos
relative to a $10^{14} \hinv\msol$ halo is larger, and the central 
galaxy red magnitudes at these mass scales are more similar.

The blends add a low-mass tail to the halo mass selection function for 
clusters of fixed optical richness.
For our found clusters with optical richness targeting $10^{14} 
\hinv\msol$ halos, we expect that $\sim 10\%$ of these systems would be 
underluminous in X--rays by a factor of two at $z=0.41$, 
growing to
$\sim 20\%$ underluminous by a factor closer to three at $z=0.99$.  The 
scatter in individual X--ray luminosities for the complete set of 
clusters is expected to be large,
$\sigma_{\ln L}  \simeq 1.2$ at high redshift, and there is 
considerable overlap in the distributions of $L_X$  expected for clean 
and blended clusters.  
 It should be noted that, observationally, high redshift low-luminosity systems
are also likely have lower signal to noise.

The galaxy number density profiles are slightly
shallower for blends than for clean clusters, and a matched spatial
filter approach may help identify and eliminate the former.  Since
some fraction of halos, those undergoing mergers especially, will also
be spatially extended, careful study of the effect of spatial
filtering on halo completeness is needed.  Alternatively, instead
of decreasing the number of blends in searches, our findings here 
suggest modeling the mass likelihood $p(M|\Ngal,z)$ as a bimodal log-normal 
distribution, with the fraction of blends, and the location and width 
of that component, included as nuisance parameters.  
This expected bimodal distribution can be incorporated into error 
estimates for cluster number counts as a function of redshift, 
for instance, along with other expected errors (such as the 5-10\% scatter
associated with red sequence associated redshifts \citealt{Gil07}).

Understanding the detailed color/magnitude trends within galaxy clusters is
key to refining red sequence cluster finding and improving its success rate.
Fortunately, data sets in hand or on the way, combined with rapidly improving
modeling methods, will lead to improvements in our understanding of high
redshift colors and their evolution.
This work will be driven largely by survey-specific mocks--- current examples
are the 2MASS \citep{Koc03}, the DEEP2 survey\citep{YanWhiCoi04}, the 2dFGRS
\citep{Eke04,Yan05} and the SDSS \citep{Mil05,Koe07,Wei06b} --- and such
efforts will be necessary for mining the rich science provided by existing
and future high redshift cluster surveys.

\medskip

We thank the anonymous referee for many helpful comments and
suggestions.  JDC thanks A. Albrecht, M. Brodwin, C. Fassnacht,
R. Gal, J. Hennawi, A. von der Linden, L. Lubin, G. De Lucia,
S. Majumdar, T. McKay, N. Padmanabhan, E. Rozo, R. Stanek and
D. Weinberg for helpful discussions and/or questions and the Galileo
Galilei Institute, the Santa Fe Cosmology Summer Workshop, and the
Aspen Center for Physics for hospitality during the course of this
work and the opportunity to present these results, and LBL for
support.  AEE thanks J. Annis and T. McKay for conversations and
acknowledges support from NASA grant NAG5-13378, from NSF ITR
ACI-0121671, and especially from the Miller Institute for Basic
Research in Science at UC, Berkeley.  MW thanks Charles Lawrence for
conversations.  DJC wishes to thank both the Aspen Center for Physics
and the Department of Physics at the University of Michigan for
hospitality, and acknowledge support from NSF grant AST507428. MW
acknowledges support from NASA and EE acknowledges support from NSF
grant AST-0206154.  The Millennium Simulation was carried out by the
Virgo Supercomputing Consortium at the Computing Centre of the Max-
Planck Society in Garching; semi-analytic galaxy catalogues are
publicly available at http://www.mpa-garching.mpg.de/Millennium/.

\section*{Appendix}

Purity and completeness are ``success rates'' used when one wants a catalogue
of a certain type of object and has obtained, via some method, a catalogue of
candidates.   A classic definition starts with the number of objects which
are both in the candidate set and in the desired target set, i.e.~the
intersection of these sets.  Dividing the number in the intersection by the
total number of target objects then gives completeness, and dividing by
the total number of candidate objects gives purity (or reliability).
These definitions go back many years in radio astronomy.  For instance
\citet{ConBalJau75} used these definitions to describe how well optical
sources were matched to radio sources as a function of search aperture radius.
In our case, target objects are halos, defined in terms of true richness
or mass, and candidate objects are clusters, defined in terms of observed
richness.

While these terms have a long history, it is not clear that such definitions
are the `single number' one wants to characterize the success of a cluster
finding algorithm.  If the properties of the sample change slowly with
e.g.~richness or mass for example, we may not wish to impose a hard threshold
on richness when computing purity.  Finding a cluster with 19 members may be
just as good as requiring 20.  We shall consider several generalizations of
the classic notions of purity and completeness below.

We note that there are several choices in all of these definitions: the two
catalogues, including their underlying data samples and the search algorithms
employed, and the definition of which clusters lie in the intersection.
Even focussing on the circular overdensity method, as we do here, there is
considerable latitude in defining both the target catalogue (e.g., specific
definition of a halo, use of halo richness or mass as the order parameter)
and the cluster candidate sample (minimum observed cluster richness, choice
of $\Delta_p$).
We consider here how purity and completeness vary with $\Delta_p$.

\begin{figure}
\begin{center}
\resizebox{3.0in}{!}{\includegraphics{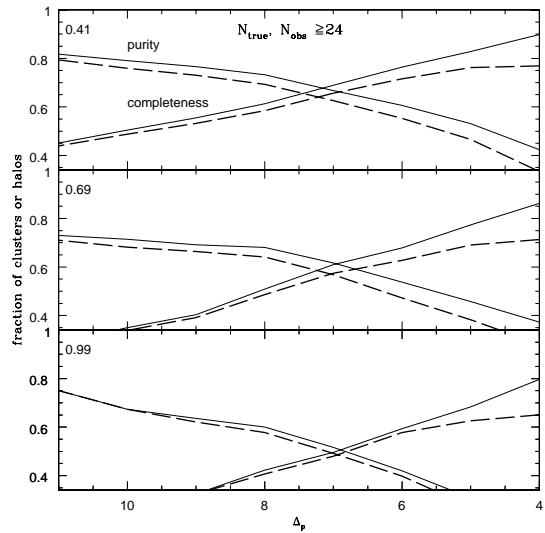}}
\end{center}
\caption{Purity and completeness as a 
function of overdensity contrast $\Delta_p$ in Eq.~\ref{eqn:codefn} for
$z=0.41$, 0.69, 0.99, top to bottom. The 
target catalogue is halos
with $24\leq N_{\rm true} \leq 100$ and the candidate clusters have
$24\leq N_{\rm obs} $. The number
of objects in the overlap of the two catalogues is divided by
the total number of objects in the target halo catalogue to get completeness
(rising lines) and by the total number of objects in the candidate cluster
catalogue to get purity (falling lines).  The overlap
of the two catalogues is either taken to be either target halos which 
contribute the most galaxies to a candidate cluster (solid line)
or only target halos which contribute at least
half of their cluster galaxies (dashed line), i.e. only halos that
match to clean clusters.}
\label{fig:true_pur_lit}
\end{figure}

Figure \ref{fig:true_pur_lit} shows the classic definitions of purity and
completeness applied to our catalogues for target halos with true richness
$24 \leq N_{\rm true} \leq 100$ and candidate clusters with
observed richness $\geq 24$.  We define a cluster and its halo to be
in the overlap of the two catalogues if the halo in the target set
contributes the most galaxies to a cluster in the candidate set.  A
more restrictive definition is to require the halo to contribute more
than half of the galaxies in a given cluster ($f_{1h} \geq 0.5$).  In
both cases, one divides the overlap number by the total number of
target halos (completeness) and total number of candidate clusters
(purity).  In Fig.~\ref{fig:true_pur_lit}, these two cases (all
$f_{1h}$ and $f_{1h}\geq 0.5$) for the overlap set are shown as smooth
and dashed lines, respectively.

We show Fig.~\ref{fig:true_pur_lit} as a function of {\it decreasing\/}
overdensity threshold, $\Delta_p$, because this mimics a search region
of increasing radial scale.  At high $\Delta_p$, purity is maximized
because the cluster sample is selecting the dense cores of the most
massive halos.  As the threshold is decreased, the enlarged search
area and lower intrinsic density within the halo increases the
frequency of best matched halos which are below the target richness threshold,
lowering the cluster sample purity.  In contrast, the completeness grows with
lower $\Delta_p$, as the increasing search scale matches and then exceeds
the radial scale used to define the halo population.
The number of observed clusters at fixed richness increases rapidly with
decreasing $\Delta_p$, improving the odds of completely matching to the
massive halo sample.  The purity and completeness curves cross at roughly
our chosen threshold, $\Delta_p\simeq 7$.

\begin{figure}
\begin{center}
\resizebox{3.0in}{!}{\includegraphics{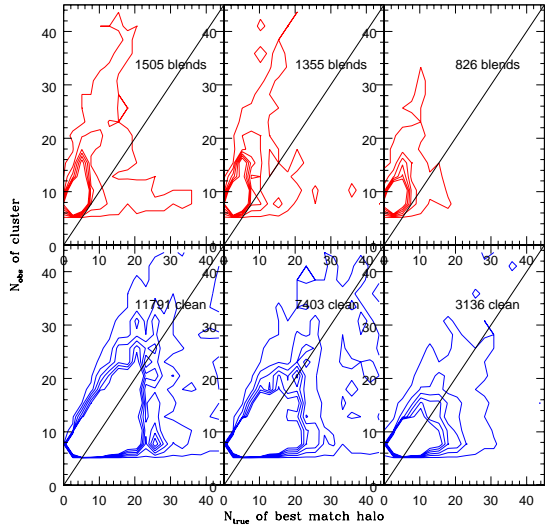}}
\end{center}
\caption{Scatter between true halo richness (x-axis) and observed cluster
richness (y-axis), for blends (top) and clean clusters (bottom), for
$z=0.41$, 0.69, 0.99 (left to right) summed over all three projection axes.
Even for the clean clusters the scatter is extremely large.
Contours differ by 10, starting at 2, and the pixel size is 2.5.
The straight line is $N_{\rm true}=N_{\rm obs}$.}
\label{fig:truefound}
\end{figure}

These measures of purity and completeness are substantially lower than unity
because, as shown in Fig.\ref{fig:truefound}, there is substantial scatter
between $N_{\rm true}$ and $N_{\rm obs}$ for matched halo-cluster pairs.  
This should not be taken as a failure of the algorithm, because the source
of impurity and incompleteness -- the scatter -- is largely understood.
There are a number of ways to take this scatter into account, and the optimal
method depends strongly on the intended use of the catalogue.  Different
target/candidate sets will vary in purity and completeness, driven by the
form of the scatter.  To characterize this one could, for example, use
$N_{\rm true} \geq N$, $N_{\rm obs} \geq N-\delta$ to calculate completeness,
and $N_{\rm true}\geq N-\delta$, $N_{\rm obs} \geq N$ to calculate purity.
If one approximates the scatter as Poisson an obvious choice for $N=25$ would be
$\delta=\sqrt{N}\simeq 5$.
For $N=24$ using $\delta=5$ raises the fractions for completeness and purity
by $\sim 0.1-0.2$ (the largest change is at $\Delta_p$ for low completeness
or purity, the smallest at high completeness or purity), while using
$\delta=10$ roughly doubles the effect.
The purity and completeness curves still cross around $\Delta_p\simeq 7$
(slightly lower for $z=0.41$)  but at a higher fraction (for $z=0.41$ 
it goes from $\sim0.7$ to $\sim0.85$ to $\sim0.95$ for shifts by 5 and 10,
respectively).
Note that the {\it a priori\/} arbitrary choice of $\delta$ should be
motivated by some understanding of how the sample properties change with
the property being used to define the sample.

\begin{figure}
\begin{center}
\resizebox{3.0in}{!}{\includegraphics{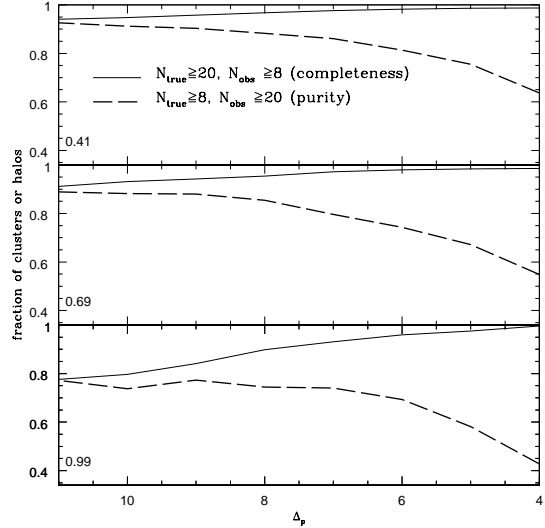}}
\end{center}
\caption{Purity and completeness as a function of overdensity contrast
$\Delta_p$ for $z=0.41$, 0.69, 0.99, top to bottom.
Unlike Fig.~\ref{fig:true_pur_lit}, the target and candidate samples used
to define purity and completeness differ.  For completeness (increasing line),
the target catalogue is halos with $20\leq N_{\rm true} \leq 100$ and the
candidate clusters are any found cluster (i.e.~$8\leq N_{\rm obs}$), both
clean and blended.  For purity (decreasing line), the candidate catalogue
comprises all clusters with $N_{\rm obs} \geq 20$, the target set is halos
with $8 \leq  N_{\rm true} \leq 100$, and a halo is only taken to be in the
intersection if it contributes the majority of galaxies to a cluster,
i.e.~this purity is simply the fraction of clean clusters, $f_{\rm clean}$.
Completeness decreases and purity increases with increasing $\Delta_p$.}
\label{fig:true_pur}
\end{figure}

Extending the candidate sample when defining completeness and the target
sample when defining purity can be taken further.  We can take the target
halo $N_{\rm true}$ above some cut and consider {\it all\/} clusters above
the minimum richness threshold (which could be as low as one) to define
completeness, for example.  In our case the minimum richness threshold is
$N_{\rm obs} \geq 8$.  The differential form of completeness for
$N_{\rm obs} \ge 8$ and $\Delta_p =7$ is shown in Fig.~\ref{fig:ffound} in
the main text.  For purity one can again reverse the limits.  If one goes to
$N_{\rm true} \geq 1$, all clusters will get matched to at least one halo,
and values close to this will have similar results.  A possibly more useful
definition of purity could be that the best matched halo contributes at least
0.5 of its partner cluster's members.   This definition of purity corresponds
to the clean fraction shown in Fig.~\ref{fig:mlike}.

Fig.~\ref{fig:true_pur} shows the purity and completeness for these less
restrictive sample definitions, taking $N=20$ and $\delta = 12$.  
The solid increasing line and the dashed
decreasing line (with decreasing $\Delta_p$) are directly analogous to their
counterparts in Fig.~\ref{fig:true_pur_lit}.  At high surface density
$\Delta_p$, the $N_{\rm obs} \geq 8$ cluster sample is incomplete with respect
to $100 \geq N_{\rm true} \geq 20$ halos because the cores of some
$N_{\rm true} \geq 20$ halos fall below the cluster richness limit of 8
members.  As mentioned in the text, high redshift high mass halos are more
likely to be disturbed than their lower redshift counterparts and thus to
fall below the overdensity threshold.  The purity of the overall cluster
sample purity is very high at $z=0.4$, but declines at higher redshift where
halo blending is more severe.

As the threshold $\Delta_p$ is lowered, the fraction of halo galaxies lying
above the projected threshold increases, but the potential for confusion by
projection also increases.  For $\Delta_p=4$, the $N_{\rm obs}\geq 8$ cluster
sample is essentially 100\% complete for $N_{\rm true} \geq 20$ halos at all
redshifts.  The overall purity of the $N_{\rm obs} \geq 20$ cluster sample is
substantially lower, dropping to values below 0.5 at $z=1$.

Another way of choosing the two samples is pursued by \citet{Roz07}, who take
Fig.~\ref{fig:truefound}, combining both blends and clean clusters, and
identify strong outliers (see their Fig.~2). They take a slice in a fixed
$N_{\rm obs}$ range to define purity (number which are not outliers over total
in slice) and a slice in a fixed $N_{\rm true}$ range to define completeness
(number which are not outliers over total in slice).
The issue of how outliers are defined and which slices in $N_{\rm obs}$,
$N_{\rm true}$ are taken will affect the detailed results, and, again, the
intended use of the catalogue needs to be taken into account before deciding
the optimal choices.

The difficulty in finding the best definition lies in trying to get two
numbers (purity and completeness) to characterize an entire joint distribution.
The full distribution of candidate properties as a function of target halo
properties is the key information required to compute expectations for an
observational catalogue.  The scatter is not a problem if its shape is
sufficiently well understood and the required accuracy for understanding this
distribution depends upon the specific use of the catalogue.
For instance, high purity (but not necessarily high completeness) might be
of interest if one is interested in high mass clusters for individual X-ray
followup, while if one wants a sample of clusters for cosmological parameters,
a scatter in mass can be included in the analysis, but high completeness is
desirable to beat down statistics.  If one can correct for scatter
perfectly, obtaining high purity is then not crucial; purity serves only to
quantify the size of the correction being applied to the data or the model.
The errors that are most important to avoid, and how well characterized our
selection function is, determines the best line of attack in the tradeoff
between purity and completeness.

\bibliographystyle{mnras}

\end{document}